\documentclass{article}
\usepackage{dsfont}
\usepackage{graphicx} 
\usepackage{amsmath}
\usepackage{subfigure}
\usepackage[subfigure,titles]{tocloft}
\usepackage[makeroom]{cancel}
\usepackage{subcaption}
\usepackage{float} 
\usepackage{enumitem} 
\usepackage[numbers]{natbib} 
\usepackage{caption} 
\usepackage[utf8]{inputenc} 
\usepackage{hyperref} 
\usepackage[toc,page]{appendix}

\title{Tutorial: VAE as an inference paradigm for neuroimaging}

\author{\makebox[\textwidth][c]{C. Vázquez-García$^{1}$, F. J. Martinez-Murcia$^{1}$, F. Segovia Román$^{1}$, Juan M. Górriz$^{1}$}}

\date{January 2025}

\begin{document}

\maketitle

\begin{center}
$^1$ Department of Signal Theory, Telematics and Communications, University of Granada, Spain
\end{center}

\begin{abstract}
    In this tutorial, we explore Variational Autoencoders (VAEs), an essential framework for unsupervised learning, particularly suited for high-dimensional datasets such as neuroimaging. By integrating deep learning with Bayesian inference, VAEs enable the generation of interpretable latent representations. This tutorial outlines the theoretical foundations of VAEs, addresses practical challenges such as convergence issues and overfitting, and discusses strategies like the reparameterization trick and hyperparameter optimization. We also highlight key applications of VAEs in neuroimaging, demonstrating their potential to uncover meaningful patterns, including those associated with neurodegenerative processes, and their broader implications for analyzing complex brain data.
\end{abstract}

\section{Introduction}
Variational Autoencoders (VAEs) have emerged as a powerful tool for unsupervised learning, offering a framework to model complex, high-dimensional data through probabilistic inference \cite{kingma2013auto, rezende2014stochastic}. Unlike traditional autoencoders, VAEs integrate principles from Bayesian inference, allowing them to generate and reconstruct data by learning latent representations that are both interpretable and continuous \cite{blundell2015weight}. This paradigm has proven particularly valuable in fields dealing with intricate and multidimensional datasets, such as neuroimaging.

Neuroimaging data, which includes structural and functional brain scans, often exhibits high dimensionality, noise, and heterogeneity. These characteristics make traditional machine learning approaches prone to overfitting or limited generalization \cite{poldrack2015long}. Moreover, the integration of neuroimaging data with other modalities—such as cognitive assessments, cerebrospinal fluid markers, or genetic information—requires robust generative models capable of capturing complex relationships while preserving interpretability. This challenge is particularly crucial when integrating multimodal datasets, such as combining structural brain scans with genetic or cognitive markers.

While the theoretical foundations of VAEs are well established, their practical implementation often comes with significant challenges. Issues such as convergence to suboptimal solutions, over-regularization leading to a loss of expressivity, or difficulties in normalizing multimodal data are common hurdles in real-world applications \cite{kingma2014adam, kucukelbir2017automatic}. Furthermore, in the context of neuroimaging, VAEs must tackle domain-specific challenges, such as the alignment of data across individuals or the scarcity of large, labeled datasets \cite{abraham2014machine}.

This tutorial aims to provide a clear and intuitive explanation of the core concepts of VAEs, providing a comprehensive yet intuitive explanation of their theoretical underpinnings. Additionally, it explores common pitfalls encountered during implementation, offering insights into strategies to mitigate these issues. Throughout the tutorial, we emphasize applications in neuroimaging, showcasing the potential of VAEs to uncover latent structures in complex brain data and enhance our understanding of neurological and psychiatric conditions. By bridging the gap between theory and practice, this tutorial serves as a guide for researchers interested in leveraging VAEs as an inference paradigm in neuroimaging.

\subsection{An opportunity: latent generative models}

Neuroimaging plays a central role in studying NDDs due to its non-invasive nature and ability to provide detailed insights into brain structure and function \cite{d2015cognitive}. However, the high dimensionality of neuroimaging data presents challenges for modeling, often leading to the curse of dimensionality, where the number of features surpasses the number of available samples \cite{fayyad1996kdd}.

The manifold hypothesis offers a potential solution, suggesting that high-dimensional data tends to concentrate around lower-dimensional latent manifolds \cite{tenenbaum2000global}. This concept has inspired a range of techniques to uncover these hidden structures, from classical approaches like PCA to advanced machine learning methods such as Autoencoders \cite{hinton2006reducing}, GANs \cite{goodfellow2014generative}, and Latent Diffusion Models \cite{rombach2022high}. For a further mathematical insight of the manifold hypothesis refer to appendix \ref{appA}.

This manifold hypothesis provides a powerful foundation for neuroimaging analysis. Neuroimaging data often lie on lower-dimensional latent manifolds, which can be leveraged to uncover meaningful patterns associated with degeneration—patterns that may not be directly observable to the naked eye-. Techniques such as Principal Component Analysis (PCA) have long been used to model neuroimaging data in clinical settings through latent variables due to its popularity and powerful capability of uncovering hidden patterns in data. Beyond PCA, a wide array of methods aim to identify the underlying manifold structure hidden within complex datasets. For example, Tenenbaum et al. \cite{tenenbaum2000global} demonstrated how data points that appear close in Euclidean space, such as those distributed along a spiral, may actually be far apart when considering the manifold's true geometry. Techniques like PCA or Multidimensional Scaling (MDS) \cite{borg2007modern} often fail to capture such complex non-linear structures, emphasizing the need for more sophisticated approaches.

Locally Linear Embedding (LLE) \cite{roweis2000nonlinear} is one such method that reconstructs the latent manifold by capturing local geometry. It represents each data point as a linear combination of its neighbors, recovering non-linear structures effectively. However, LLE has limitations: it is sensitive to noise, which can distort the manifold, and relies on sufficient data density to ensure accurate connections between neighboring points. In scenarios where data is sparse or noisy, its performance may degrade significantly.

Recent advances in Deep Neural Networks (DNNs) have introduced a new paradigm for representing data in non-linear latent spaces. Models like Autoencoders (AEs), Generative Adversarial Networks (GANs), and Latent Diffusion Models (LDMs) offer powerful tools for uncovering and representing latent structures. But, why favor deep learning over simpler methods? Unlike traditional techniques, which rely on predefined mathematical procedures, DNNs are highly flexible and adaptable, capable of learning complex patterns across diverse data types \cite{lecun2015deep, he2016deep}. Their ability to model non-linearity is particularly critical, as real-world data rarely conforms to linear assumptions.

Moreover, DNNs draw inspiration from the brain, mimicking processes of neural encoding. The brain processes an immense volume of high-dimensional sensory input by encoding it into lower-dimensional, abstract representations. This principle is well-documented in vision research. For instance, Rowekamp and Sharpee \cite{rowekamp2017cross} analyzed the response of V2 neurons in the visual cortex to natural stimuli. These neurons encode complex visual inputs from earlier cortical stages into simpler structures, such as multi-edge features, which help identify contrasts and object boundaries. Subsequent stages of neural processing abstract these features further, transforming them into complex ideas—a process still under active investigation in psychology.

\section{The variational autoencoder as a latent generator paradigm}

\subsection{The theory of the Variational Autoencoder}

When working with 2D images or 3D volumes, one of the most commonly used frameworks of DNN is the Variational Autoencoder (VAE), which is a modification of the usual Autoencoder, based on Bayesian inference. While Autoencoders learn a fixed latent representation, VAEs learn a probabilistic distribution of the latent representations. This model was introduced by Kingma in its work \cite{kingma2013auto}, where he proposed a way to learn the intractable posterior distribution that encode the real data into the latent representation. Imagine we have a dataset $X$ of i.i.d. samples of a continuous variable. We assume that there is some unknown random hidden process that generates the data, concerning some unobserved variable $z$. This unobserved variable is drawn from some prior distribution $p(z)$. Therefore, since every sample from $X$ is generated by the hidden process, each value $x$ is generated by the conditional distribution $p(x|z)$. For simplicity, we assume that both the prior function and the likelihood $p(x|z)$ come from a parametric family of functions that we parameterise as $\theta$ and that their PDFs are differentiable with respect to $z$ and $\theta$. The general idea is to find the parameter $\theta$ that maximizes the likelihood $p(x|z)$. In summary, we just want to find which parameters make it more believable that our real data $x$ come from the latent representation $z$. This problem is modelled by the joint distribution of $X$ and $Z$, known as the generative model, given by:
\begin{equation}\label{joint_dist}
    p(x,z) = p(x)p(z|x) = p(z)p(x|z).
\end{equation}
$p(x)$ is the marginal likelihood of $x$, integrated over every possibility of $z$, and $p(z|x)$ is the true posterior density, that is, how probable is the value of $z$ given the observation $x$. In order to encode the data and obtain the latent representation we need to find the true posterior density, which tells us how data is encoded. If we isolate it from \eqref{joint_dist} we obtain the expression:
\begin{equation}\label{true_posterior}
    p(z|x) = \frac{p(z)p(x|z)}{p(x)}.
\end{equation}
However, this is intractable in every case that $p(x)$ is intractable, which is every practical case. It is not hard to image due to the high dimensionality that the integral $p(x) = \int p(z) p(x|z) dz$ is intractable. In order to address this issue, Kingma introduced a variational inference model to approximate the true posterior density with an inference posterior $q_{\phi}(z|x)$, where $\phi$ is a parameter. In variational autoencoders, we substitute the intractable true posterior \eqref{true_posterior} with this approximating inference model $q_{\phi}$ parameterized with a deep neural network known as encoder. The mathematical formalism of the VAE allows us to obtain a lower bound for the log-likelihood. If we assume i.i.d. data points, then we can write the marginal log-likelihood of a single observation as:
\begin{equation}
    logp(x) = log\left( \frac{p(z)p(x|z)}{p(z|x)} \right) = -logp(z|x) + \underbrace{log(p(z)p(x|z))}_{=log(p(x,z))}.
\end{equation}
Introducing the approximating inference model $logq_\phi(z|x)$ by adding and substracting in one side of the equation, we get:
$$
logp(x) =  -logp(z|x) + log(p(x,z)) + log\left( \frac{q_{\phi}(z|x)}{q_{\phi}(z|x)} \right) = 
$$
$$
= log\frac{q_{\phi}(z|x)}{p(z|x)} + logp(x,z) - logq_{\phi}(z|x).
$$
Now, let us take the expectation with respect to $q_{\phi}(z|x)$ of both sites:
\begin{equation}\label{log likelihood divergence}
logp(x) = \underbrace{E_{q_{\phi}}\left[ log\frac{q_{\phi}(z|x)}{p(z|x)} \right]}_{D_{KL}(q||p)} + E_{\phi}\left[ logp(x,z) - logq_{\phi}(z|x) \right].
\end{equation}
Take into account that $p(x)$ does not depend on $q_{\phi}$ and thus its expectation is its own value $p(x)$. The first term on the right is the definition of the Kullback-Leibler divergence \cite{kullback1951information}, which is a measure of the difference between two distributions. The KL divergence is always non negative and reaches its lowest value $0$ when both distributions are equal. Therefore, we can write a lower bound for the log likelihood as:
\begin{equation}\label{ELBO}
    log p(x) \geq E_{\phi}\left[ logp(x,z) - log q_{\phi}(z|x) \right] =: \mathcal{L}(\theta,\phi;x),
\end{equation}
which is known as the Evidence Lower Bound Objective. Notice that the distribution $p$ is chosen from a parametric family paramererized by $\theta$, as previously mentioned, and so the ELBO depends on both $\theta$ and $\phi$. It is worth mentioning that the gap between the ELBO and the real log likelihood of the data is exactly the KL divergence. However, we can not directly use \eqref{ELBO} since we do not know the joint distribution $p(x,z)$. For practical use we can rewrite the previous expression as:
\begin{equation}\label{ELBO compute}
    log p(x) \geq E_{\phi}[log p(x|z)] - D_{KL}(q_{\phi}(z|x)||p(z)),
\end{equation}
which now depends on the prior distribution. It is common to use a Normal distribution $\mathcal{N}(0,1)$ as a prior distribution, for simplicity.

\subsection{The practical implementation of the Variational Autoencoder and its complications}

Now that we have seen the theory, it is appropriate to talk about its practical implementation and its use in neuroimaging. So far, we have talked about the generation of latent manifolds, but we have also mentioned that VAE has a true posterior density $p(z|x)$ that accounts for the encoding of the real data in its latent representation, and a likelihood $p(x|z)$ that accounts for the decoding of the latent manifold into real-like reconstructed data. So, the architecture of the VAE consists of both an Encoder $E_{\phi}$ and a Decoder $D_{\theta}$. If we look at the ELBO function \eqref{ELBO compute} we see that it consists simply of a reconstruction term $E_{\phi}[logp(x|z)]$ that tells us how well the decoder reconstructs the data from the latent space under the inference model $\phi$, and a regularization term  $D_{KL}(q_{\phi}(z|x)||p(z))$ that tells us how close the inference model is to the prior distribution. Hence, the loss of the model can be easily computed by computationally implementing both terms. For the reconstruction term one can choose a reconstruction function that best fits the needs of its problem. A simple and efficient choice is the mean squared error (MSE), which provides physical meaning. However, in the context of images, this reconstruction measure is not well suited to match perceived visual quality. The mean squared error cannot retrieve the structural features of the images because it does not take into account the spatial relationships throughout the image. In addition, it tends to focus on the data points that perform the worst \cite{girod1993s, eskicioglu1995image, wang2002universal}. For neuroimaging, a good choice is the Structural Dissimilarity Index (DSSIM) between two images or volumes, which is based on the assumption that vision in humans is adapted to perceive structural information from its surroundings. The DSSIM is computed as $1-$SSMI, where SSMI is the Structural Similarity Index introduced in the work \cite{wang2004image}. The idea is to use a measure that can objectively quantify the perceived quality of the image by a human. SSMI not only takes into account the pixel-wise difference like MSE does, but also accounts for the spatial relationships of the pixels, which allows it to retrieve the structural information with precision. 

On the right-hand side of the ELBO \eqref{ELBO compute} we have the divergence loss, which accounts for the regularization of the latent space. In the case of a Normal prior distribution $p(z)\sim \mathcal{N}(0,1)$ and a Gaussian inference model $q_{\phi}(z|x)\sim \mathcal{N}(\mu(x), \sigma(x))$, the expression of the KL divergence is:
\begin{equation}\label{KL div}
    D_{KL}(\mathcal{N}(\mu(x), \sigma(x))||\mathcal{N}(0,1)) = \frac{1}{2}\sum_{i=1}^{d}(1+log(\sigma_i^2(x)) - \mu_i(x)^2 - \sigma_i^2(x)),
\end{equation}
where $\mu_i(x)$ and $\sigma^2_i(x)$ are the mean and variance of the $i-th$ variable of the latent space.
The next step is to choose an architecture for our VAE. For neuroimaging, the natural choice is the C-VAE, a VAE consisting of convolutional layers \cite{lecun2015deep}.
\begin{figure}[!ht] 
    \centering 
    \includegraphics[width=0.8\textwidth]{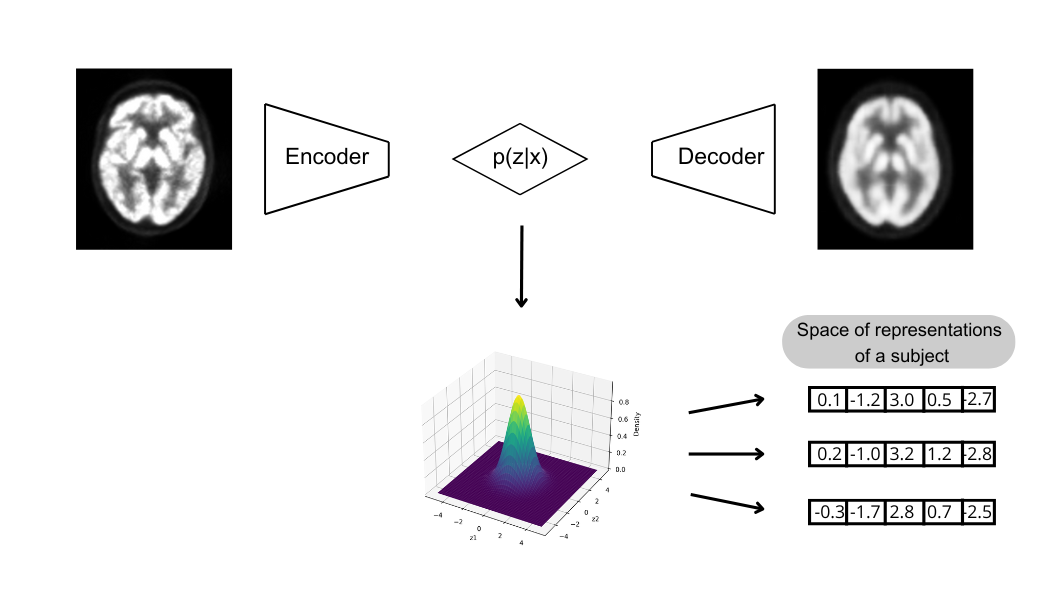}  
    \caption{Conceptual scheme of the VAE. A neural network codifies the information into a lower dimensional manifold characterized by a probability distribution $p(z|x)$ which can be sampled to analyze the latent data. A neural network is also trained to decode the latent information back to the data space} 
    \label{VAE_scheme}  
\end{figure}
A convolution is a mathematical function that measures the degree of superposition of two functions and produces an output function. In the context of signal processing or imaging, convolutions are widely used to extract features like borders, textures or even patterns from the input. For convolutional layers, we have both the input image and a kernel, with lower dimension than the input. By convolutioning them, the kernel is able to extract patterns from the image into lower dimensional representations known as feature maps. Since convolutions are linear functions, we apply a non-linear activation function such as Rectified Linear Unit (ReLU) to introduce non-linearity in the model. This allows the network to learn more complex representations of the data. The following work \cite{shridhar2019comprehensive} gives a detailed guide to work with CNN in the context of Variational Inference. CNNs are extensively used in various fields, such as facial recognition, object detection, imaging segmentation or analysis of medical images among others, due to their powerful feature extraction \cite{schroff2015facenet, redmon2016you, ronneberger2015u, litjens2017survey}.

\subsubsection{The gradient problem and the reparameterization trick}

Once the loss function computation is clear, all that is left is to backpropagate this loss through our network in order to update the weights. To do so, we need to compute the gradient of the ELBO \eqref{ELBO compute} with respect to the parameters of the model $\theta$, $\phi$, which is a bit tricky. On the first hand, the gradient with respect to $\theta$ is:
\begin{equation}
    \nabla_{\theta}\mathcal{L} = \nabla_{\theta}\mathds{E}_{q_{\phi}(z|x)}[logp(x|z)] - \nabla_{\theta}D_{KL}(q_{\phi}(z|x)||p(z)).
\end{equation}
To compute the first term we can replace the expectation value with an unbiased Monte Carlo estimator \cite{robert1999monte}:
$$
\mathds{E}_{q(z|x)}\left[\log p(x|z)\right] \approx \frac{1}{N} \sum_{i=1}^{N} \log p(x|z_i),
$$
which allows us to rewrite the gradient as:
\begin{equation}\label{gradient_theta}
\nabla_{\theta} \mathcal{L} \approx \frac{1}{N} \sum_{i=1}^{N} \nabla_{\theta} \log p(x|z_i),
\end{equation}
where the term of the divergence is zero because it does not depend on the parameter $\theta$. If we now take the gradient with respect to $\phi$ we have the following expression:
\begin{equation}\label{gradient_phi}
    \nabla_{\phi}\mathcal{L} \approx  \frac{1}{N} \sum_{i=1}^{N} \nabla_{\phi} \log p(x|z_i) - \nabla_{\phi}D_{KL}(q_{\phi}(z|x)||p(z)).
\end{equation}
The problem with the first term of the expression \eqref{gradient_phi} is that, if we sample $z$ directly from $q_{\phi}(z|x)\sim \mathcal{N}(\mu_{\phi}(x), \sigma_{\phi}(x))$, the relationship between $z$ and $\phi$ is not smooth due to the random nature of the sampling, and thus it is not differentiable. On the other hand, since $q_{\phi}(z|x)$ is differentiable with respect to $\phi$ there is no issue with the gradient of the regularisation term. The solution to this issue, as proposed by Kingma \cite{kingma2019introduction}, is straightforward. Given that the problem arises from the stochastic nature of the sampling process, it can be addressed by drawing a sample from a standard normal distribution and subsequently reparameterizing it as:
$$
z = \mu(\phi;x) + \sigma(\phi;x)\cdot\epsilon,
$$
where $\epsilon\sim\mathcal{N}(0,1)$. Since the distribution of the random variable $\epsilon$ is independent of every parameter or $x$, then the reparameterization $z = f(\phi, x, \epsilon)$ is now differentiable with respect to the parameters. This transformation is known as the reparameterization trick.  The idea behind the reparameterization trick is that since $z$ is calculated using $\mu(x;\phi)$ and $\sigma(x;\phi)$, which are smooth functions with respect to $\phi$, small variations of $\phi$ (which occur during the gradient) will lead to small variations of $z$ unless $\epsilon$ takes extreme values, which is unlikely to happen because we control it with a normal distribution. Without the reparameterization trick, direct sampling of $z$ from $q_{\phi}(z|x)$ is unpredictable even when $\phi$ varies slowly, and may introduce large variations.

With everything set, we can use our deep learning environment of choice and implement our network. Pythoch is a common framework due to its large variety of functions to implement deep neural networks and its usage of tensors, which is fundamental to work with GPUs \cite{paszke2019pytorch}. We just have to implement some convolutional layers, feed the neuroimages as input, encode using the previous variational formalism, obtain our latent representation, and finally decode using the generative model.

However, as expected, even though the theory seems to be straightforward, there are many things to take into account in practice. For instance, if we want a very powerful network that is able to extract every single detail, we may think that we should add as many layers as we desire to add complexity to the model. However, we know that neural networks are universal function approximators \cite{hornik1989multilayer}, which means that if the network is too complex, it might memorize every single property of the training set that may not be well suited for our problem, i.e., it becomes prone to overfitting data \cite{geman1992neural}.

\subsubsection{The information preference problem and the hyperparameter space}

Aside from overfitting, there is a more concerning issue for variational autoencoders, regarding its autoencoding capability. Even though VAEs are interpreted as autoencoders, meaning that the generated data is close to the real data, the conditions in which this happens are not discussed in the original article \cite{kingma2013auto}. As discussed in the article \cite{chen2016variational}, VAEs do not always autoencode and they may not even use the information in the latent space unless the decoder is weak. In this work they use a Bits-Back coding approach to prove that if the decoder $p(x|z)$ is able to reconstruct the data without using the information of the latent space, then it will not use it, in which case it would set $q_{\phi}(z|x)$ to simply be equal to the prior $p(z)$. This means that we have to achieve a balance between the complexity of the model, which could lead to both overfitting and loss of latent information, and a significant latent representation, in order to obtain accurate patterns of neuroimaging. This issue occurs because, as we have seen in the ELBO formula \eqref{ELBO}, the KL divergence simply does not care about the information of the latent space, it only cares about the similarity between the variational model and the prior. We may think that if $q_{\phi}(z|x)$ is close to $p(z)$ that would be good, because we are minimizing loss. However, if our $q_{\phi}(z|x)$ is too close to the prior, we would not capture the specific inter-subject variability and thus our representation would not be meaningful. To understand this issue, recall that our context is neuroimaging. Imagine we want to construct a VAE that is able to capture patterns of neurodegeneration of a certain neurodegenerative disease, in order to apply statistical techniques to model the development of the disease. If our network is flexible enough it could reconstruct the neuroimages without the information of the latent space, where the patterns of neurodegeneration for each subject are located. If we decoded using our flexible decoder, we would obtain images that are very similar to the input images, but our latent representation would carry no information. 

 We need to bear in mind that the goal is not to obtain a very accurate reconstruction of the images but to find a meaningful representation. We only need to capture the patterns and the inter-subject variability, we do not need to reconstruct every detail that may not carry any information. To deal with this problem we can use a different divergence in place of the Kullback-Leibler divergence. Researchers show in the work \cite{zhao2017infovae} that under some assumptions, the ELBO can be rewritten as:
\begin{equation}\label{info ELBO}
    \mathcal{L}_{\text{Info-VAE}} = \mathds{E}_{q_{\phi}(z|x)}[log p(x|z)] - \lambda D(q_{\phi}(z|x)||p(z)),
\end{equation}
with the constraint that $\lambda\rightarrow\infty$. In practice, it is sufficient that $\lambda$ is at least of the same order as the log likelihood. In this work they also propose the Maximum-Mean Discrepancy (MMD) measure, which is a metric that computes the distance between two distributions, by comparing all of their moments. Using this divergence instead of KL will maximize the mutual information between the input data $x$ and the latent variables $z$, and thus the information of the latent space will be meaningful and used by the decoder.

\begin{figure}[ht]
    \centering
    \begin{subfigure}
        \centering
        \includegraphics[angle=90, width=0.6\textwidth]{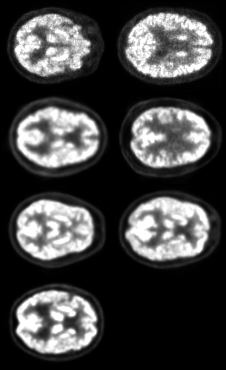}
        \label{fig:imagen1}
    \end{subfigure}
    \vspace{1em} 
    \begin{subfigure}
        \centering
        \includegraphics[angle=90, width=0.6\textwidth]{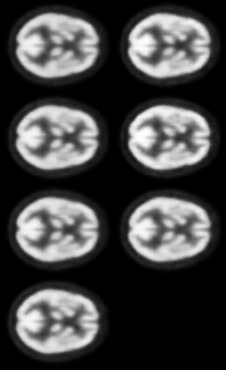}
        \label{fig:imagen2}
    \end{subfigure}
    \caption{On top: Input volume slices from the neuroimaging database. On bottom: reconstructed images from random subjects using a variational autoencoder using the volumes on top as input. In the figure we see that the reconstruction of each subject is the same. This happens because the model is stuck at the mean local minimum, and thus it is not able to capture the inter-subject variability.}
    \label{mean_brain_kl}
\end{figure}

Having addressed the information preference problem, we can resume our analysis of the latent generative model to derive meaningful representations of our subjects. We feed our database to the VAE model, obtain the latent representation of neuroimaging for the neurodegenerative disease we are working with, and finally decode and plot curves of loss to see the performance of our model. But, to our surprise, when we take a look at our reconstructed images or volumes we find the following image (\ref{mean_brain_kl}). 

After all the issues we have already dealt with and solved, something has gone wrong again. But, what exactly happened? We are encountering a common issue in deep learning: our model appears to be trapped in a local minimum of the loss function. The loss function is a non-linear multi-variate function that lives in a space of large dimensionality. Its structure is highly complex and it has many local minima \cite{zhang2021understanding, sutskever2013importance, tishby2015deep}. Among those minima, there is one local attractor, which is the mean of the distribution. In the ELBO, one simple solution for $q_{\phi}(z|x)$ to be close to the prior $p(z)$ is simply to approximate $\mu(x)$ to $0$ and $\sigma(x)$ to $1$, since the prior is a normal distribution. This is catastrophic, because in this case we are not capturing the variability. Instead, what our model is doing is compute and decode an average brain of the subjects of the database, paying no attention to the inter-subject variability. This is exactly what we are seeing in the figure (\ref{mean_brain_kl}).

In this simple but illustrative guide, the author \cite{gopubby2023} provides a good explanation about the problem of local minima in functions. In order to avoid being trapped in a local minimum we need to find the appropriate parameters such that our model will evolve into a global minimum or, at least, a lower local minimum. However, as we have already mentioned, the number of parameters in a loss function is extremely large. We have to take into account the parameters $\theta$ and $\phi$, but we also have the parameters of each convolutional and fully connected layer, which can be of the order of tens of thousands and we do not have direct control over them. We also have the hyperparameters, such as the batch size, the regularization weight $\lambda$ \cite{higgins2017beta}, the learning rate, or the number of epochs. To avoid local minima, we have to explore the space of them and adjust them accordingly until we find a combination of hyperparameters that yields significative results. 

\subsection{Latent representations as a benchmark for analysis in neuroimaging}

Variational Autoencoders are explicit probabilistic models, meaning they learn an explicit distribution over the latent space, which is typically designed to be accessible and interpretable, opposed to other latent generative models such as GAN or LDM, which learn implicit representations that are useful for performing different tasks such as synthesis or reconstruction, but that do not provide great insights of the latent variables.

Now that our VAE model is finally operative, we can perform some statistical analysis to these representations. Remember that all the information that we have extracted is encoded within our latent space. To exploit this latent variables there are many statistical techniques that we can apply. Firstly, we know that our model is correctly reconstructing volumes and inter-subject variability by using the latent variables, which means that there is information within those variables that understand the structural differences between subjects. For example, there may be some variables that encode the size of the ventricles, or the thalamus or even the deterioration of the temporal lobes. We are interested in which properties are codified within each latent variable. However, this relationship may not be simple. In the case of structural size, we may find a variable that is able to codify it, but if we try to find a variable that codifies dementia we would probably find nothing. Instead of a single latent variable, it could be codified as a linear (or non-linear) combination of the variables of the latent space. If we were able to find such a relationship we would have a model that is able to codify and understand the patterns related to dementia. 

For instance, imagine we want to see whether dementia is encoded in our latent variables, and we want to identify which variables are more relevant for such encoding. One way is to use an SVR model, which learns a function to predict the target variable (e.g., dementia diagnosis) based on the latent variables. SVR uses kernels that map the input data into higher-dimensional spaces where non-linear relationships can become linear. However, this transformation is performed implicitly, meaning that there is no explicit formula showing how the latent variables combine to predict the target, which makes the model act like a black box. This is not desirable since our main goal is to interpret the information of the latent space.

To address this issue, alternative methods, such as the Generalized Linear Model (GLM), can be employed. GLMs extend linear regression by introducing a link function that connects the expected value of the response variable to a linear combination of the latent variables. This allows the model to capture non-linear relationships through explicit transformations. Unlike SVR, where the kernel implicitly transforms the data, GLMs offer direct control over the link function, enabling interpretability. While GLMs may be less flexible than SVR in terms of the transformations they can model, they provide valuable insights into how the latent variables contribute to the target variable.

Other techniques, such as polynomial or spline regression, can also capture linear and non-linear relationships between the latent variables and the response variable, offering additional flexibility for statistical modeling.

Even though most research in the literature focuses on latent representations for reconstruction or data synthesis, recent studies have explored their use in extracting meaningful insights. Below, we highlight notable works that leverage latent representations in neuroimaging:

\begin{enumerate}
    \item \textbf{Fusing Functional and Structural Neuroimaging Data}:  
    In \cite{geenjaar2021fusing}, the authors employ a VAE model to integrate functional and structural neuroimaging data into a shared latent space using a unified encoder-decoder scheme. This approach enables the model to learn cross-modal patterns, facilitating interpolation and interpretability between modalities. By integrating information from both modalities, the study demonstrates improved insights compared to analyzing each modality independently.  

    \item \textbf{Cross-Modality Latent Representations}:  
    The work in \cite{vazquez2024cross} develops a joint VAE model combining neuroimaging and clinical data. The authors demonstrate that specific latent variables can reconstruct clinical data (e.g., UPDRS scores) using neuroimaging representations, and vice versa \ref{UPDRS}. This is achieved through a cross-modality loss function that enforces shared latent representations during training. The study highlights the potential of such models to capture cross-modal patterns, particularly in the context of Parkinson’s disease, where the latent representations of medical volumes and clinical data collectively recover the UPDRS score, a key measure of symptom severity and progression.  

    \begin{figure}[ht]
        \centering
        \includegraphics[width=0.8\textwidth]{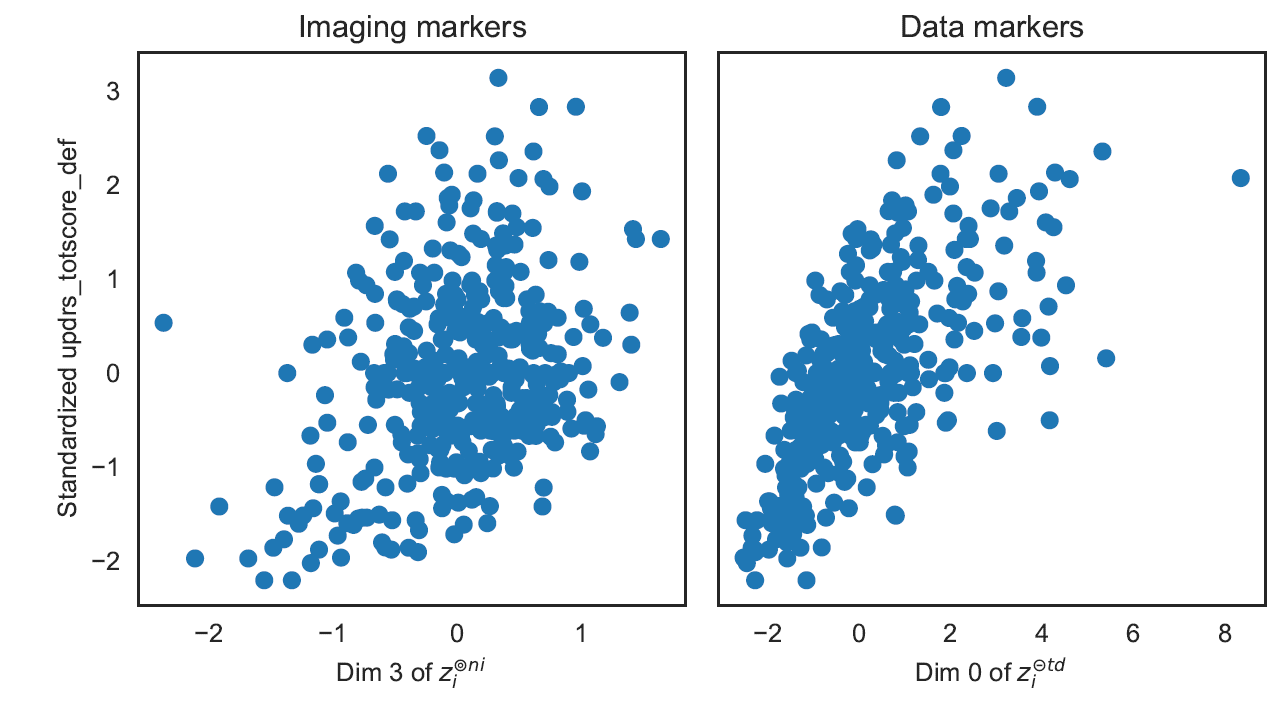}
        \caption{UPDRS score vs. values of latent variables. Certain latent variables exhibit a correlation with the UPDRS score, reflecting Parkinson's severity in the subject.}
        \label{UPDRS}
    \end{figure}  

    \item \textbf{Latent Representations for Longitudinal Patterns}:  
    In \cite{qiang2020deep}, the authors apply the VAE formalism to 4D fMRI data to uncover longitudinal patterns of functional activity and derive functional brain networks. This work showcases the utility of VAEs in capturing dynamic changes in brain activity over time.  

    \item \textbf{Predicting Brain Age Trajectories}:  
    Another study, \cite{chadebec2022image}, leverages VAE-derived latent representations to compute brain trajectories through linear modeling. By analyzing these trajectories, the authors successfully predict age progression and assess the impact of specific latent variables on these predictions.  
\end{enumerate}  

These works illustrate how latent representations extend beyond reconstruction tasks, offering valuable insights into complex neuroimaging patterns and their relationships to clinical or functional outcomes. They emphasize the interpretability and versatility of VAE models in uncovering meaningful information from high-dimensional neuroimaging data.

\appendix
\section*{Appendix A: Mathematical Implications of the Concentration Phenomenon}
\addcontentsline{toc}{section}{Appendix A: Mathematical Implications of the Concentration Phenomenon}
\refstepcounter{section} 
\label{appA}

To better understand why high-dimensional data often concentrates around lower-dimensional structures, we can analyze the geometry of high-dimensional spaces. Consider the volume of a hypersphere, a generalization of a circle ($2D$) or a sphere ($3D$) to n-dimensional space. The volume of such a hypersphere with radius $R$ is given by:

\begin{equation}\label{volSphere}
    V_n(R) = \frac{\pi^{n/2}R^n}{\Gamma\left(\frac{n}{2}+1\right)},
\end{equation}
where $\Gamma$ denotes the Gamma function, a generalization of the factorial to real and complex numbers. While this formula captures the growth of volume in low dimensions, as $n$ increases, an interesting phenomenom emerges: the volume of the hypersphere becomes increasingly concentrated near its surface. This counterintuitive result implies that, in high dimensions, most points are pushed toward the surface, leaving the center sparsely populated. Consequently, most of the space appears "empty", with the volume concentrating near the surface. To explore this, we calculate the volume of a thin shell near the surface of the hypersphere. The volume of the shell is given by:
\begin{equation}
    V_{shell} = V_n(R) - V_n(R-\epsilon),
\end{equation}
where $R\in\mathds{R}^{n+1}$ is the radius of the hypersphere and $\epsilon<<1$ is a small scalar. Using \eqref{volSphere} we can write:
\begin{equation}
    V_{shell} = V_n(R) - \frac{\pi^{n/2}(R-\epsilon)^n}{\Gamma\left(\frac{n}{2}+1\right)}.
\end{equation}
Using Taylor approximation around $R$ we can write $(R-\epsilon)^n\approx R^n - nR^{n-1}\epsilon$, and therefore the volume of the shell is:
\begin{equation}
    V_{shell} \approx \cancel{V_n(R)} - \cancel{\frac{\pi^{n/2}(R)^n}{\Gamma\left(\frac{n}{2}+1\right)}} + n\frac{\pi^{n/2}R^{n-1}\epsilon}{\Gamma\left(\frac{n}{2}+1\right)},
\end{equation}
$$
V_{shell} \approx n\frac{\pi^{n/2}R^{n-1}\epsilon}{\Gamma\left(\frac{n}{2}+1\right)} .
$$
Now, we can compute the fraction of the total volume that the shell represents: 
$$ratio_{shell} = \frac{V_{shell}}{V_n(R)} \approx \frac{n\epsilon}{R}.$$
This result demonstrates that, for a fixed $\epsilon$, the fraction of volume in the shell near the surface increases with $n$. In fact, even as $\epsilon\to 0$, the ratio approaches $1$ as $n\to\infty$. This implies that, in high dimensions, nearly all the volume is concentrated in a thin shell adjacent to the surface. 

To gain further intuition, consider the equation for the radius of a hypersphere:
\begin{equation}\label{radius}
    r_n = x_1^2 + x_2^2 + \dots +x_n^2 + x_{n+1}^2,
\end{equation}
where the sphere is centered at the origin. If we choose a random point and calculate the probability of it being close to the center, we find that this probability decreases as the number of dimensions increases. This is because the points have exponentially more configurations in higher dimensions, pushing them away from the center.

In summary, in high-dimensional spaces, points tend to cluster close to a lower-dimensional manifold.  This phenomenon is a manifestation of the concentration of measure phenomenon \cite{ledoux2001concentration, donoho2000high}. Note, however, that this is a mathematical abstraction, and latent manifolds of real-world data are often more complex.

\bibliographystyle{unsrt} 

\bibliography{references}

\end{document}